\documentclass[letterpaper,journal]{IEEEtran} 
\usepackage{amsmath,amsfonts}
\usepackage{algorithmic}
\usepackage[ruled,linesnumbered]{algorithm2e}
\usepackage{array}
\usepackage[caption=false,font=normalsize,labelfont=sf,textfont=sf]{subfig}
\usepackage{textcomp}
\usepackage{stfloats}
\usepackage{url}
\usepackage{balance}
\usepackage{verbatim}
\usepackage{graphicx}
\usepackage[colorlinks]{hyperref}
\hypersetup{citecolor=blue}
\usepackage{threeparttable}
\usepackage{multirow}
\usepackage{amssymb}
\usepackage{bm} 
\usepackage{geometry}
\newcommand{\E}{\mathbb{E}}
\newcommand{\C}{\mathbb{C}}

\newcommand{\herm}{H}
\newcommand{\trans}{T}

\newtheorem{proposition}{\bfseries Proposition}
\newtheorem{remark}{\bfseries Remark}

\usepackage[font=small,skip=2pt]{caption}

\usepackage{enumitem}
\setlist{nosep}

\begin{document}

\title{\LARGE Finite-Aperture Fluid Antenna Array Design: Analysis and Algorithm}
	
\author{Zhentian Zhang,~\IEEEmembership{Graduate Student Member,~IEEE}, 
            Kai-Kit Wong,~\IEEEmembership{Fellow,~IEEE}, 
            Hao Jiang,~\IEEEmembership{Senior Member,~IEEE}, 
            Farshad Rostami Ghadi,~\IEEEmembership{Member,~IEEE}, 
            Hyundong Shin,~\IEEEmembership{Fellow,~IEEE}, and 
            Yangyang Zhang
\vspace{-8mm}

\thanks{Z. Zhang and H. Jiang are with the National Mobile Communications Research Laboratory, Southeast University, Nanjing, 210096, China and H.~Jiang is also with the School of Artificial Intelligence, Nanjing University of Information Science and Technology, Nanjing 210044, China. (e-mail: zhentianzhangzzt@gmail.com, jianghao@nuist.edu.cn).}

\thanks{K. K. Wong and F. R. Ghadi are with the Department of Electronic and Electrical Engineering, University College London, Torrington Place, WC1E 7JE, United Kingdom  (e-mails: \{kai-kit.wong, f.rostamighadi\}@ucl.ac.uk). K. K. Wong is also affiliated with the Department of Electronic Engineering, Kyung Hee University, Yongin-si, Gyeonggi-do 17104, Republic of Korea.}
\thanks{H. Shin is with the Department of Electronics and Information Convergence Engineering, Kyung Hee University, Yongin-si, Gyeonggi-do 17104, Republic of Korea (e-mail: hshin@khu.ac.kr).}
\thanks{Y. Zhang is with Kuang-Chi Science Limited, Hong Kong SAR, China (e-mail: yangyang.zhang@kuang-chi.com).}

\thanks{Corresponding authors: K.-K. Wong}
}

\maketitle

\begin{abstract}
Finite-aperture constraints render array design nontrivial and can undermine the effectiveness of classical sparse geometries. This letter provides universal guidance for fluid antenna array (FAA) design under a fixed aperture. We derive a closed-form Cram\'er--Rao bound (CRB) that unifies conventional and reconfigurable arrays by explicitly linking the Fisher information to the geometric variance of port locations. We further obtain a closed-form probability density function of the minimum spacing under random FAA placement, which yields a principled lower bound for the minimum-spacing constraint. Building upon these analytical insights, we then propose a gradient-based algorithm to optimize continuous port locations. Utilizing a simple gradient update design, the optimized FAA can achieve about a $30\%$ CRB reduction and a $42.5\%$ reduction in mean-squared error.
\end{abstract}


\vspace{-2mm}
\section{Introduction}
\IEEEPARstart{L}{inear} array design \cite{VanTrees2002} is pivotal for sensing estimation such as angle-of-arrival (AoA). Port placement strongly greatly affects estimation precision and side-lobe behavior. {\em If the antenna aperture budget is flexible}, various array structures \cite{Liu2017}, e.g., nested arrays, coprime arrays, and minimum redundancy arrays (MRA) \cite{Moffet1968}, can significantly outperform the half-wavelength uniform linear array (ULA). However, array design becomes more challenging under {\em finite-aperture} constraints, in which existing methods are not necessarily effective when the aperture is limited, and a flexible architecture is preferred.

Recently, the fluid antenna system (FAS) was proposed \cite{Wong2020} and it is a hardware-agnostic system concept that regards the antenna as a reconfigurable physical-layer resource to broaden system design with new degree of freedom (DoF) \cite{new2024tutorial,hong2025contemporary,new2025flar,Wutuo2026}. FAS has been employed to utilize antenna reconfiguration for flexible beamforming \cite{JXU_beam}, sensing \cite{tang2025fullduplex,Zhang2025} and multiple access \cite{FAS2}. Despite the potential, fluid antenna array (FAA) design under finite-aperture constraints is not explored. While \cite{Zhang2025} provided an upper bound for finite-aperture AoA sensing, fundamental limits and design rules for FAA are missing.

In this work, we provide universal guidance for FAA design under finite-aperture budgets by clarifying fundamental geometric limits under fixed array length, and we further propose a practical FAA design algorithm to validate the analysis. The contributions are summarized as follows:
\begin{itemize}
\item First, we derive a closed-form Cram\'er--Rao Bound (CRB) that captures different geometries under finite-aperture constraints, unifying reconfigurable array analyses.
\item We characterize the minimum port spacing from a distributional perspective by deriving a closed-form probability density function (PDF), matching Monte-Carlo results.
\item Finally, we propose a gradient-based algorithm based on the derived insights, validating the analytical results.
\end{itemize}

{\em Notations}---Scalars, vectors, and matrices are denoted by $a$, $\bm a$, and $\bm A$, respectively. $(\cdot)^\trans$, $(\cdot)^\herm$, $(\cdot)^*$, $\|\cdot\|_2$, $\otimes$, and $\mathrm{vec}(\cdot)$ are transpose, Hermitian transpose, complex conjugate, $\ell_2$-norm, Kronecker product, and vectorization. $\mathbb{E}[\cdot]$ is expectation.

\vspace{-2mm}
\section{System Model}
\subsection{Finite Aperture Configuration}
Consider a linear FAA at the base station (BS) with $M$ active ports whose locations are to be optimized. The  physical aperture normalized by the wavelength $\lambda$ is $W_{\max}=(M-1)/2$, which matches a ULA of $M$ elements spaced by $\lambda/2$. Exploiting the flexibility of FAA, the port locations are modeled as continuous variables collected in $\bm p=[p_1,\dots,p_M]^\trans\in\mathbb{R}^M$, where $p_m$ denotes the normalized position of the $m$-th port. To maximize the effective aperture, the {\em first} and {\em last} ports are fixed at the aperture boundaries, i.e., $p_1 = 0$, and $p_M = W_{\max}$. Simultaneously, the intermediate ports are constrained within the same aperture, i.e., $0 < p_2 < \dots < p_{M-1} < W_{\max}$. Additionally, to avoid the overlapping of multiple ports, a minimum inter-port spacing $d_{\min}$ is prescribed as
\begin{equation}\label{eq:dmin_constraint}
p_m - p_{m-1} \ge d_{\min},~m=2,\dots,M.
\end{equation}

\vspace{-2mm}
\subsection{Signal Model}
Without loss of generality, we consider line-of-sight transmission. The steering vector $\bm{a}(\theta, \bm{p}) \in \C^M$ of AoA $\theta$ under port placement $\boldsymbol{p}$ can be expressed as
\begin{equation}\label{eq:steer}
\bm{a}(\theta, \bm{p}) = \left[ e^{-j 2\pi p_1 \cos \theta}, \dots, e^{-j 2\pi p_M \cos \theta} \right]^\trans.
\end{equation}
The received signal vector $\mathbf{y}(t) \in \mathbb{C}^{M}$ in far-field transmission at the $t$-th time snapshot is modeled as
\begin{equation}\label{eq:y_t}
\mathbf{y}(t) = \bm{a}(\theta, \bm{p}) s(t) + \mathbf{n}(t),
\end{equation}
where $s(t)$ is the transmitted source signal with average power $P_s = \E[|s(t)|^2]$ and $\mathbf{n}(t) \sim \mathcal{CN}(\mathbf{0}, \sigma_n^2 \mathbf{I}_M)$ represents the additive white Gaussian noise (AWGN) with zero mean and variance $\sigma_n^2$. The signal-to-noise ratio (SNR) at the transmitter side is defined as $\text{SNR} \triangleq \frac{P_s}{\sigma_n^2}$.

\vspace{-2mm}
\subsection{Virtual Array Transformation}
To align with \cite{Zhang2025}, the second-order statistics of the received signal is exploited to expand the sensing aperture by virtual array transformation from the covariance matrix. Define the snapshot matrix as $\bm{Y} \triangleq [\bm{y}(1),\bm{y}(2),\dots,\bm{y}(T)] \in \C^{M\times T}$ and the corresponding covariance is given by $\bm{R}_{\bm{Y}} \triangleq \E[\bm{y}(t)\bm{y}(t)^\herm]$. Consider the independence among AWGN and signal component, the covariance matrix $\bm{R}_{\bm{Y}}$ is approximated as
\begin{equation}\label{eq:cov_approx}
\bm{R}_{\bm{Y}} \approx P_s \bm{a}(\theta, \bm{p})\bm{a}(\theta, \bm{p})^\herm + \sigma_n^2 \bm{I}_M,
\end{equation}
which can be readily extended into the case of multiple AoAs if the multi-path components are deemed as uncorrelated and the cross-terms equal to zeros in the covariance matrix \cite{VanTrees2002}. Then vectorizing $\bm{R}_{\bm{Y}}$ yields the virtual received signal
\begin{equation}\label{eq:vec_cov}
\bm{y}_v = \text{vec}(\bm{R}_{\bm{Y}}) \approx  P_s\big(\bm a(\theta,\bm p)^*\otimes \bm a(\theta,\bm p)\big) + \sigma_n^2\,\text{vec}(\bm I_M),
\end{equation}
where the Kronecker term is known as the virtual difference co-array (DCA) and is determined by the set of position differences $\{p_u - p_v\}$. Conveniently, one could form a sensing AoA codebook by discretizing the angular domain into a grid $\Theta = \{\vartheta_1, \dots, \vartheta_N\}$. The sensing codebook $\bm{A}(\bm{p}) \in \C^{M^2 \times N}$ is then constructed, whose $n$-th column is $\bm{c}_n(\bm{p}) = \bm{a}(\vartheta_n, \bm{p})^* \otimes \bm{a}(\vartheta_n, \bm{p})$, where the entry corresponding to the antenna pair $(u, v)$ is $[\bm{c}_n(\bm{p})]_{(v-1)M+u} = e^{-j 2\pi (p_u - p_v) \cos \vartheta_n}$.

\vspace{-2mm}
\section{Minimum Spacing Analysis}\label{subsec:spacing_analysis}
The minimum spacing in \eqref{eq:dmin_constraint} is essential to FAA to prevent port overlap. However, the statistical behavior of port spacing under finite-aperture placement remains unclear. This section rigorously justifies the necessity of the minimum-spacing constraint $d_{\min}$ by analyzing port spacing statistics under a random placement strategy. Specifically, $M$ active ports are selected independently and uniformly at random within the fixed aperture $[0, W_{\max}]$. Let $X_{(1)} < X_{(2)} < \dots < X_{(M)}$ denote the order statistics representing the sorted physical positions of active ports. The joint PDF is uniform over the correspondingly simplex $\mathcal{D} = \{ (x_1, \dots, x_M) \mid 0 \le x_1 < \dots < x_M \le W_{\max} \}$, given by $f(\mathbf{x}) = M!/W_{\max}^M$. Let $\Delta_{\min} = \min_{1 \le i \le M-1} \{ X_{(i+1)} - X_{(i)} \}$ denote the minimum spacing random variable and we derive its PDF below.

\begin{proposition}\label{prop.1}
The expected minimum spacing for FAA should be set no smaller than
\begin{equation}\label{eq:E_Delta_min}
\mathbb{E}[\Delta_{\min}] = \int_0^{\frac{W_{\max}}{M-1}} \delta f_{\Delta_{\min}}(\delta) d\delta = \frac{W_{\max}}{M^2 - 1},
\end{equation}
where $\delta \in [0, \frac{W_{\max}}{M-1}]$ and the PDF is given by
\begin{equation}\label{eq:pdf_min_spacing}
f_{\Delta_{\min}}(\delta) = \frac{M(M-1)}{W_{\max}} \left( 1 - \frac{(M-1)\delta}{W_{\max}} \right)^{M-1}.
\end{equation}
\end{proposition}

\begin{IEEEproof}
Initially, let the event $\mathcal{E} = \{ \Delta_{\min} > \delta \}$ imply that the spacing between every pair of adjacent ordered elements must exceed $\delta$, i.e., $X_{(i+1)} - X_{(i)} > \delta$ for all $i=1, \dots, M-1$. Then, we derive the complementary cumulative distribution function (CCDF), denoted as $\mathbb{P}(\Delta_{\min} > \delta)$, in a {\em geometric model of probabilstic manner}. However, calculating the {\em volume} of this constrained high-dimensional region directly is intractable due to the coupled boundary conditions. To loosen the calculation demand, we introduce a new set of variables $\mathbf{Y} = [Y_1, \dots, Y_M]^T$ defined as
\begin{equation}\label{eq.trans}
Y_k = X_{(k)} - (k-1)\delta.
\end{equation}
Therefore, the difference between adjacent $Y_k$ is
\begin{align}\label{eq.transfer}
Y_{i+1} - Y_i &= \left( X_{(i+1)} - i\delta \right) - \left( X_{(i)} - (i-1)\delta \right) \nonumber \\
&= (X_{(i+1)} - X_{(i)}) - \delta.
\end{align}
By \eqref{eq.transfer}, the original constraint $X_{(i+1)} - X_{(i)} > \delta$ is equivalently converted to the standard ordering constraint $Y_{i+1} > Y_i$. Simultaneously, the original variable domain $X_{(M)} \le W_{\max}$ transforms to $Y_M + (M-1)\delta \le W_{\max} \rightarrow Y_M \le W_{\max} - (M-1)\delta$. Let $W' = W_{\max} - (M-1)\delta$ denote the \textit{effective aperture}. 

The minimum-spacing constraint induces multiple exclusion zones in the joint domain, yielding a fragmented admissible region with coupled boundaries. The linear transformations in \eqref{eq.trans} and \eqref{eq.transfer} remove these zones by index-dependent shifts of the ordered coordinates, mapping the region bijectively and volume-preservingly onto a contiguous standard simplex with reduced aperture. Thus, the feasible volume is obtained from the simplex $\mathcal{S}' = \{ (y_1,\dots,y_M)\mid 0 \le y_1 < \cdots < y_M \le W' \}$.

The Jacobian of this linear transformation is the determinant of a triangular matrix with ones on the diagonal, i.e., $|\det(\mathbf{J})| = 1$. Therefore, the volume calculation preserves the differential relationship of $d\boldsymbol{x} = d\boldsymbol{y}$. The volume of the standard simplex $\mathcal{S}'$ is simply $(W')^M / M!$. The probability is the ratio of the valid volume to the total volume ($W_{\max}^M / M!$)
\begin{multline}
\mathbb{P}(\Delta_{\min} > \delta) = \frac{\text{Vol}(\mathcal{S}')}{\text{Vol}(\mathcal{D})} \\
= \frac{(W_{\max} - (M-1)\delta)^M}{W_{\max}^M}=\left( 1 - \frac{(M-1)\delta}{W_{\max}} \right)^M.
\end{multline}
Differentiating $F_{\Delta_{\min}}(\delta) = 1 - \mathbb{P}(\Delta_{\min} > \delta)$ with respect to $\delta$ gives the PDF of $\Delta_{\min}$ in (\ref{eq:pdf_min_spacing}), defined for the support $\delta \in [0, \frac{W_{\max}}{M-1}]$. The upper bound of the support, i.e., $\delta_{\max} = \frac{W_{\max}}{M-1}$, corresponds to the ULA configuration. Since the ULA maximizes the minimum spacing by distributing elements evenly, no randomized placement can achieve a minimum spacing larger than this value.
\end{IEEEproof}

\begin{remark}
{\em This result reveals a fundamental limitation of random FAA. While the element spacing in a ULA decreases linearly with $M$, i.e., $\propto M^{-1}$, the expected minimum spacing in a random FAA decays \textit{quadratically}, i.e., $\propto M^{-2}$.}
\end{remark}

\vspace{-2mm}
\section{Universal CRB for Finite-Aperture Array}\label{sec:theory}
In this section, we derive the exact closed-form expression for the CRB under finite-aperture budget. By explicitly {\em relating the CRB to the antenna position variance}, we provide a theoretical benchmark universally applicable for different systems and different antenna placement. Recalling the received signal in \eqref{eq:y_t}, to generalize the CRB to different carrier frequency, we rewrite the steering vector $\mathbf{a}(\theta,\boldsymbol{p})$ in \eqref{eq:steer} as
\begin{equation}
\mathbf{a}(\theta,\boldsymbol{p}) = \left[ e^{-j k p_1 \cos(\theta)}, \dots, e^{-j k p_M \cos(\theta)} \right]^T,
\end{equation}
where $k = 2\pi/\lambda$ is the wavenumber. If the antenna positions are normalized by $\lambda$, i.e., $p_m \leftarrow p_m/\lambda$, then the steering component becomes $e^{-j2\pi p_m\cos\theta}$, equivalent to setting $\lambda=1$. In the sequel, we will omit the variable term $\boldsymbol{p}$ and treat it as deterministic constants to derive CRB for angle $\theta$.

Under the deterministic signal model, the Fisher Information Matrix (FIM) element $J_{\theta\theta}$ with respect to the angle $\theta$ is given by the Slepian-Bangs formula \cite[Section~3.3]{Stoica2005}
\begin{equation}\label{eq:fim_general}
J_{\theta\theta} = \frac{2 T P_s}{\sigma_n^2} \text{Re} \left\{ \dot{\mathbf{a}}^H(\theta) \mathbf{P}_{\mathbf{a}}^{\perp} \dot{\mathbf{a}}(\theta) \right\},
\end{equation}
where $\text{Re} \left\{\cdot\right\}$ takes the real part of the variable, $P_s$ is the signal power, $\dot{\mathbf{a}}(\theta)$ denotes the partial derivative $\frac{\partial \mathbf{a}(\theta)}{\partial \theta}$, and $\mathbf{P}_{\mathbf{a}}^{\perp}$ is the orthogonal projection matrix onto the null space of $\mathbf{a}(\theta)$:
\begin{equation}\label{eq:Pa_general}
\mathbf{P}_{\mathbf{a}}^{\perp} = \mathbf{I}_M - \frac{\mathbf{a}(\theta)\mathbf{a}^H(\theta)}{\mathbf{a}^H(\theta)\mathbf{a}(\theta)}=\mathbf{I}_M - \frac{1}{M}\mathbf{a}(\theta)\mathbf{a}^H(\theta),
\end{equation}
where the simplification considers $|[\mathbf a]_m|=1$ and $\mathbf a^H{\bf a}=M$. Next, we explain how to derive the CRB step-by-step below.

\subsection{Derivative of the Steering Vector}
To obtain an explicit relationship between estimation accuracy and array geometry (port placement $\boldsymbol{p}$), the quadratic term in \eqref{eq:fim_general}. Since the $m$-th element of the steering vector is $[\mathbf{a}(\theta)]_m = e^{-j k p_m \cos(\theta)}$, its derivative with respect to $\theta$ is
\begin{multline}\label{eq:a_dot}
\frac{\partial [\mathbf{a}]_m}{\partial \theta} = \frac{\partial}{\partial \theta} \left( e^{-j k p_m \cos(\theta)} \right) \\
= e^{-j k p_m \cos(\theta)} \cdot (-j k p_m) \cdot (-\sin(\theta))
= j k p_m \sin(\theta) [\mathbf{a}]_m,
\end{multline}
which can be written in a more compact form $\dot{\mathbf{a}}(\theta) = j k \sin(\theta) \mathbf{D} \mathbf{a}(\theta)$, where $\mathbf{D} = \text{diag}(p_1, p_2, \dots, p_M)$ is a diagonal matrix containing the antenna positions.

\subsection{Expansion of the Quadratic Form}
Plugging \eqref{eq:a_dot} into the quadratic term $\mathcal{Q} = \dot{\mathbf{a}}^H \mathbf{P}_{\mathbf{a}}^{\perp} \dot{\mathbf{a}}$ in \eqref{eq:fim_general}
\begin{equation}\label{eq:Q_terms}
\mathcal{Q} = \dot{\mathbf{a}}^H \left( \mathbf{I}_M - \frac{1}{M}\mathbf{a}\mathbf{a}^H \right) \dot{\mathbf{a}}
= \underbrace{\dot{\mathbf{a}}^H \dot{\mathbf{a}}}_{\text{Term I}} - \frac{1}{M} \underbrace{\dot{\mathbf{a}}^H \mathbf{a}\mathbf{a}^H \dot{\mathbf{a}}}_{\text{Term II}},
\end{equation}
where \textit{Term I} and \textit{Term II} can be further simplified. For {\em Term I}, plugging \eqref{eq:a_dot} and recalling that $|[\mathbf{a}]_m|=1$, the squared norm of the derivative vector can be written as
\begin{align}
\dot{\mathbf{a}}^H \dot{\mathbf{a}} = \sum_{m=1}^M \left| j k p_m \sin(\theta) [\mathbf{a}]_m \right|^2= k^2 \sin^2(\theta) \sum_{m=1}^M p_m^2.
\end{align}
Then for {\em Term II}, we first evaluate the inner product 
\begin{align}\label{eq.TermI}
\mathbf{a}^H \dot{\mathbf{a}} &= \sum_{m=1}^M [\mathbf{a}]_m^* \cdot \left( j k p_m \sin(\theta) [\mathbf{a}]_m \right) \nonumber \\
&= j k \sin(\theta) \sum_{m=1}^M \underbrace{[\mathbf{a}]_m^* [\mathbf{a}]_m}_{=1} p_m = j k \sin(\theta) \sum_{m=1}^M p_m.
\end{align}
Consequently, Term II becomes
\begin{align}\label{eq.TermII}
\dot{\mathbf{a}}^H \mathbf{a}\mathbf{a}^H \dot{\mathbf{a}} = \left| \mathbf{a}^H \dot{\mathbf{a}} \right|^2= k^2 \sin^2(\theta) \left( \sum_{m=1}^M p_m \right)^2.
\end{align}
Substituting \eqref{eq.TermI} and \eqref{eq.TermII} back into \eqref{eq:Q_terms} yields
\begin{equation}
\mathcal{Q} = k^2 \sin^2(\theta) \left[ \sum_{m=1}^M p_m^2 - \frac{1}{M} \left( \sum_{m=1}^M p_m \right)^2 \right].
\label{eq:Q_combined}
\end{equation}

Let $\bar{p} = \frac{1}{M} \sum_{m=1}^M p_m$ denote the geometric centroid of the array. Utilizing the variance identity $\sum (x_i - \bar{x})^2 = \sum x_i^2 - \frac{1}{M}(\sum x_i)^2$, the term in the brackets of \eqref{eq:Q_combined} is exactly the \textit{effective aperture variance}, denoted as
\begin{equation}
\sum_{m=1}^M p_m^2 - \frac{1}{M} \left( \sum_{m=1}^M p_m \right)^2 = \sum_{m=1}^M (p_m - \bar{p})^2 \triangleq \mathcal{L}_{\text{geo}}(\mathbf{p}).
\end{equation}
From \eqref{eq:Q_combined}, we have the closed-form quadratic term
\begin{equation}
\mathcal{Q}\triangleq \dot{\mathbf{a}}^H(\theta)\mathbf{P}_{\mathbf{a}}^{\perp}\dot{\mathbf{a}}(\theta)=k^2\sin^2(\theta)\,
\mathcal{L}_{\text{geo}}(\mathbf{p}),
\label{eq:Q_final_again}
\end{equation}
where $
		\mathcal{L}_{\text{geo}}(\mathbf{p})
		\triangleq
		\sum_{m=1}^{M}(p_m-\bar p)^2$, $
		\bar p=\frac{1}{M}\sum_{m=1}^{M}p_m$. Noting that $\mathcal{Q}$ is real and non-negative, we hence have $\mathrm{Re}\{\mathcal{Q}\}=\mathcal{Q}$. Substituting \eqref{eq:Q_final_again} into \eqref{eq:fim_general}, we obtain
	\begin{align}
		J_{\theta\theta}
		&=
		\frac{2TP_s}{\sigma_n^2}\,
		\mathrm{Re}\left\{
		\dot{\mathbf{a}}^H(\theta)\mathbf{P}_{\mathbf{a}}^{\perp}\dot{\mathbf{a}}(\theta)
		\right\}=
		\frac{2TP_s}{\sigma_n^2}\,
		\mathcal{Q}
		\nonumber\\
		&=
		\frac{2TP_s}{\sigma_n^2}\,
		k^2\sin^2(\theta)\,
		\mathcal{L}_{\text{geo}}(\mathbf{p}).
		\label{eq:FIM_after_sub}
	\end{align}
Using the SNR definition, \eqref{eq:FIM_after_sub} becomes $J_{\theta\theta}
	=
	2T\cdot \mathrm{SNR}\cdot
	k^2\sin^2(\theta)\,
	\mathcal{L}_{\text{geo}}(\mathbf{p})$. Finally, we derive the universal closed-form CRB via the inverse of the FIM element \cite[Chapter~3]{Kay1993}
	\begin{align}
		\mathrm{CRB}(\theta)
		=
		\left(J_{\theta\theta}\right)^{-1}=
		\frac{1}{2T\cdot \mathrm{SNR}\cdot
			k^2\sin^2(\theta)\,
			\mathcal{L}_{\text{geo}}(\mathbf{p})}.
		\label{eq:CRLB_k_form}
	\end{align}
	\begin{remark}
		{\em If we use the physical wavenumber $k=2\pi/\lambda$,
		\begin{align}
			\mathrm{CRB}(\theta)
			&=
			\frac{1}{2T\cdot \mathrm{SNR}\cdot
				\left(\frac{4\pi^2}{\lambda^2}\right)\sin^2(\theta)\,
				\mathcal{L}_{\text{geo}}(\mathbf{p})}
			\nonumber\\
			&=
			\frac{\lambda^2}{8\pi^2 T\cdot \mathrm{SNR}\cdot
				\sin^2(\theta)\,
				\mathcal{L}_{\text{geo}}(\mathbf{p})}.
			\label{eq:crlb_final}
		\end{align}
		For the wavelength-normalized case in \eqref{eq:steer}, the factor $\lambda^2$ in \eqref{eq:crlb_final} is absorbed into
		$\mathcal{L}_{\text{geo}}(\mathbf{p})$, and the CRB can be written as
		$
		\mathrm{CRB}(\theta)=\frac{1}{8\pi^2 T\cdot \mathrm{SNR}\cdot \sin^2(\theta)\cdot \mathcal{L}_{\text{geo}}(\mathbf{p})}
		$ with $\lambda=1$.}
\end{remark}

\begin{remark}\label{remark3}
{\em The expression in \eqref{eq:crlb_final} reveals a trade-off between \textbf{estimation precision} and \textbf{ambiguity suppression}. Maximizing the geometric variance $\mathcal{L}_{\text{geo}}$ by placing elements near the boundaries ($p_m\to 0$ or $W_{\max}$) minimizes the CRB and achieves the highest precision, but induces \textbf{false spectral peaks} and severe beam misdirection. In contrast, uniform spacing suppresses these false peaks at the cost of a \textbf{smaller effective aperture}, resulting in lower precision.}
\end{remark}

\vspace{-2mm}
\section{Array Placement Algorithm Design}\label{sec.algorithm_design}
Remark~\ref{remark3} has thoroughly explained the trade-off between the main-lobe (estimation precision) and the side-lobe (ambiguity suppression) under finite-aperture budget. Another problem is how to conduct practical design under the observations in \eqref{eq:CRLB_k_form}. This section provides an example for practical design starting from the following remark.

\begin{remark}\label{remark.4}
{\em In \cite[(35)]{Zhang2025}, an AoA mean-squared error (AoAMSE) upperboud is predicted as
\begin{equation}
\mathrm{AoAMSE}=\mathbb{E}\left\{\left|\cos\theta_i-\cos\hat{\theta}_i\right|^2\right\}\leq\frac{16\sigma_z^4\gamma_{\max}(\boldsymbol{p})}{\bar{\lambda}^2(\boldsymbol{p})M^4},
\end{equation}
where $\sigma_z^2$ denotes the variance of the interference-plus-noise component, $\gamma_{\max}(\boldsymbol{p})$ is the largest eigenvalue of $\boldsymbol{A}^{H}(\boldsymbol{p})\boldsymbol{A}(\boldsymbol{p})$ where $\boldsymbol{A}(\boldsymbol{p})$ is the AoA sensing matrix determined by port placement and $\bar{\lambda}^2(\boldsymbol{p}) = \frac{1}{M^2} \sum_{u,v} (p_u - p_v)^2$.}
\end{remark}

Remark~\ref{remark.4} encourages us to formulate the antenna placement into an optimization problem of $\arg\min_{\boldsymbol{p}}\gamma_{\max}(\boldsymbol{p})/\bar{\lambda}^2(\boldsymbol{p})$. With logarithm-conversion, we get the AoAMSE minimization-optimization expression as
\begin{equation}\label{eq.Mini_J}
\mathop{\arg\min}\limits_{\boldsymbol{p}}J(\boldsymbol{p}) = \ln(\gamma_{\max}(\boldsymbol{p})) - \ln(\bar{\lambda}^2(\boldsymbol{p})),
\end{equation}
where if the Lagrange's identity is applied, we observe that the term $\bar{\lambda}^2(\mathbf{p}) = \frac{1}{M^2} \sum_{u,v} (p_u - p_v)^2$ is directly proportional to the geometric variance, i.e., $\sum_{u=1}^M \sum_{v=1}^M (p_u - p_v)^2 = 2M \sum_{m=1}^M (p_m - \bar{p})^2 = 2M \mathcal{L}_{\text{geo}}(\boldsymbol{p})$.
	Thus, minimizing $-\ln(\bar{\lambda}^2)$ becomes equivalent to minimizing the CRB, while minimizing $\ln(\gamma_{\max})$ suppresses sidelobes. Expression \eqref{eq.Mini_J} thereby seeks the Pareto-optimal configuration between {\em estimation precision} and {\em ambiguity resolution}. In the sequel, we propose a gradient-based algorithm to solve \eqref{eq.Mini_J}.

\subsection{Gradient Target Function Formulation}
Here, we first clarify the Gram matrix and its dominant eigenvalue as $\bm{Q}(\bm p) \triangleq \bm{A}(\bm{p})^\herm \bm{A}(\bm{p})$, $	\gamma_{\max}(\bm p) \triangleq \lambda_{\max}(\bm Q(\bm p))$ and specify the effective squared aperture term as $	\bar{\lambda}^2(\bm{p}) \triangleq \frac{1}{M^2} \sum_{u=1}^M \sum_{v=1}^M \left( 2\pi (p_u - p_v) \right)^2$. Based on \eqref{eq.Mini_J}, the following target function is formulated as
\begin{subequations}
\begin{align}
\min_{\bm{p}} &~\ln \left( \lambda_{\max} (\bm{A}(\bm{p})^\herm \bm{A}(\bm{p})) \right) - \ln \left( \frac{1}{M^2} \sum_{u,v} (2\pi(p_u - p_v))^2 \right) \\
\text{s.t.} &~p_1 = 0,~p_M = W_{\max},~p_{m} - p_{m-1} \ge d_{\min}.
\end{align}
\label{eq:opt_problem}
\end{subequations}

\subsection{Gradient Update Derivation}
To apply gradient descent, we derive the gradient of $J(\bm{p})$ with respect to the $m$-th position $p_m,m=2,\dots,M-1$, which consists of two terms
	\begin{equation}
		\frac{\partial J}{\partial p_m} = \frac{1}{\gamma_{\max}} \frac{\partial \gamma_{\max}}{\partial p_m} - \frac{1}{\bar{\lambda}^2} \frac{\partial \bar{\lambda}^2}{\partial p_m},
		\label{eq:gradJ_two_terms}
	\end{equation}
	where the two-derivative terms are:
	\begin{equation}\label{eq:grad}
		\begin{aligned}
				\frac{\partial \gamma_{\max}}{\partial p_m} = \bm{u}_{\max}^\herm \frac{\partial \bm{Q}}{\partial p_m} \bm{u}_{\max},~
				\frac{\partial \bar{\lambda}^2}{\partial p_m} = \frac{16\pi^2}{M^2} \sum_{k=1}^M (p_m - p_k),
		\end{aligned}
	\end{equation}
	where $\bm u_{\max}$ is the unit-norm eigenvector of $\bm Q$ associated with $\gamma_{\max}$. In \eqref{eq:gradJ_two_terms}, the term $-\frac{\partial \bar{\lambda}^2}{\partial p_m}$ acts like pushing the antenna elements away from the array's centroid to maximize spatial spread, while $\frac{\partial \gamma_{\max}}{\partial p_m}$ adjusts positions to minimize correlation.

\subsection{Algorithm Design}
To solve \eqref{eq:opt_problem}, a projected gradient descent (PGD) algorithm is proposed in Algorithm~\ref{alg:pgd}. The key modification is that we only update $p_2, \dots, p_{M-1}$. The {\em projection} step is conducted by a feasibility restoration mapping: sorting, edge pinning, and enforcing minimum spacing through forward/backward corrections, where $\beta$ and $\alpha$ are the momentum coefficient and gradient coefficient respectively. Notably, gradient-method is initialization-sensitive but guarantees to a local minimum value \cite[Section~2.3]{Bertsekas1999}. We initialize the placement vector $\boldsymbol{p}$ by \cite[Table~II]{Moffet1968} or \cite[Table~II]{Zhang2025} for good performance.

\subsection{Computational Complexity}
The computational cost is dominated by the construction and eigen-decomposition of the Gram matrix $\bm{Q} \in \mathbb{C}^{N \times N}$. Specifically, the per-iteration complexity scales as $\mathcal{O}(N^3 + M^2 N^2)$, where $\mathcal{O}(N^3)$ accounts for the eigenvalue evaluation and $\mathcal{O}(M^2 N^2)$ for the matrix construction and gradient updates.

\begin{algorithm}[t!]
\caption{Joint Continuous Port Optimization}\label{alg:pgd}
{\small
		\SetAlgoLined
		\KwIn{$M$, Aperture $W_{\max}$, Angle Grid $\Theta$, Step $\alpha$, Min spacing $d_{\min}$.}
		\KwOut{Optimized positions $\bm{p}^*$.}
		
		\textbf{Initialization:} 
		Get discrete MRA indices $\mathcal{I}_{\text{MRA}}$ \cite[Table~II]{Moffet1968} or \cite[Table~II]{Zhang2025}\;
		Scale indices: $\bm{p}^{(0)} = \frac{\mathcal{I}_{\text{MRA}}}{\max(\mathcal{I}_{\text{MRA}})} \times W_{\max}$\;
		
		\For{$t = 0$ \KwTo $T_{\max}$}{
			Compute $\gamma_{\max}$ and $\bar{\lambda}^2$\;
			Compute $\nabla \gamma_{\max}$ and $\nabla \bar{\lambda}^2$ via \eqref{eq:grad}\;
			Combine gradients: $\nabla J = \frac{\nabla \gamma_{\max}}{\gamma_{\max}} - \frac{\nabla \bar{\lambda}^2}{\bar{\lambda}^2}$\;
			\textbf{Update:} $\tilde{p}_m = \beta p_m^{(t)} - \alpha \nabla J_m$\;
			\textbf{Projection:} \\
			1. Sort $\tilde{\bm{p}}$\;
			2. Pin Edges: Set $\tilde{p}_1 = 0, \tilde{p}_M = W_{\max}$\;
			3. Enforce $p_m - p_{m-1} \ge d_{\min}$ (e.g., forward/backward corrections)\;
			\If{Converged}{Break\;}
}}
\end{algorithm}

\vspace{-2mm}
\section{Numerical Results}
Here, we evaluate our analytical results and algorithm design. The common setups are as follows. All antenna positions are normalized by the wavelength ($\lambda=1$), and the aperture length is set to $W_{\max}=(M-1)\times \frac{1}{2}$, corresponding to a half-wavelength ULA. For codebook construction and evaluations of all $\gamma_{\max}$ and $\bar{\lambda}^2$, we adopt a uniform grid $\Theta$ with $\theta\in[0,\pi]$ with $N=180$ grid points. Different schemes are compared, and their antenna placement methods are summarized as follows:
\begin{itemize}
\item \textit{ULA:} $p_m=(m-1)\times \frac{1}{2}$.
\item \textit{Discrete FAS (Scaled MRA):} the MRA indices are linearly scaled into $[0,W_{\max}]$ via $p=\frac{\mathcal I}{\max(\mathcal I)}W_{\max}$.
\item \textit{Continuous FAS:} initialized by the scaled MRA positions and optimized using Algorithm~\ref{alg:pgd}.
\end{itemize}

The antenna placement set of {\em Discrete FAS} is initialized following \cite[Table~II]{Zhang2025}. The minimum spacing is chosen according to the random-FAS expectation derived in Section~\ref{subsec:spacing_analysis}, i.e., $d_{\min}=\frac{W_{\max}}{M^2-1}$. For Algorithm~\ref{alg:pgd}, we set $\alpha=5\times 10^{-4}$, $\beta=0.9$, and $T_{\max}=1000$. All sensing codebooks are constructed from the covariance matrix.

\begin{figure}[t]
\centering
\subfloat[]{\includegraphics[width=0.495\linewidth]{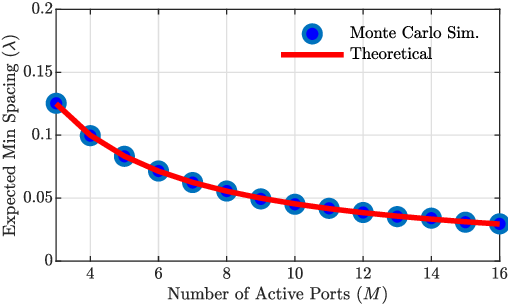}}\hspace{0pt}%
\subfloat[]{\includegraphics[width=0.495\linewidth]{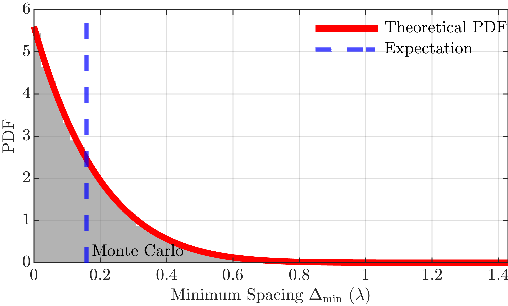}}%
\caption{Empirical and theoretical $\mathbb E[\Delta_{\min}]$ of the minimum placement constraint $d_{\min}$: {\em (a)} $M\in\{3,4,\dots,16\}$, $W_{\max}=(M-1)\times 0.5$; {\em (b)} $M=8$, $W=10$.}\label{fig:d_min}
\end{figure}

\begin{figure}[t!]
\centering
\includegraphics[width=\columnwidth]{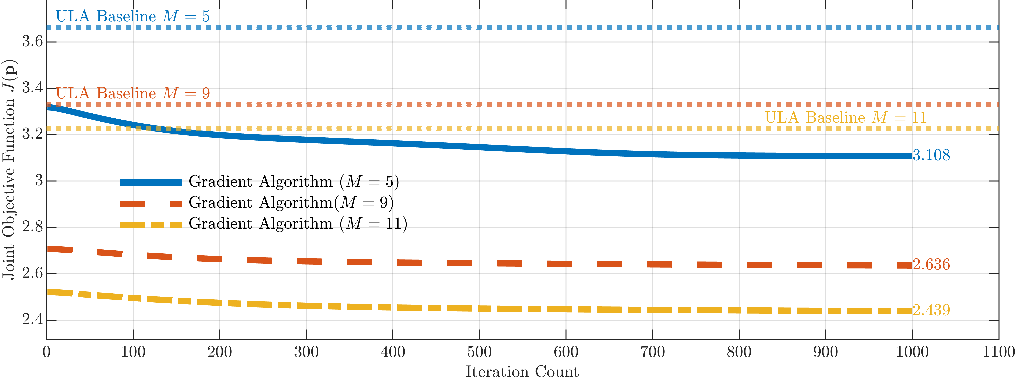}
\caption{Convergence behavior of Algorithm~\ref{alg:pgd} under $M\in\{5,9,11\}$.}\label{fig:convergence}
\end{figure}

\begin{figure}[t!]
\centering
\includegraphics[width=\columnwidth]{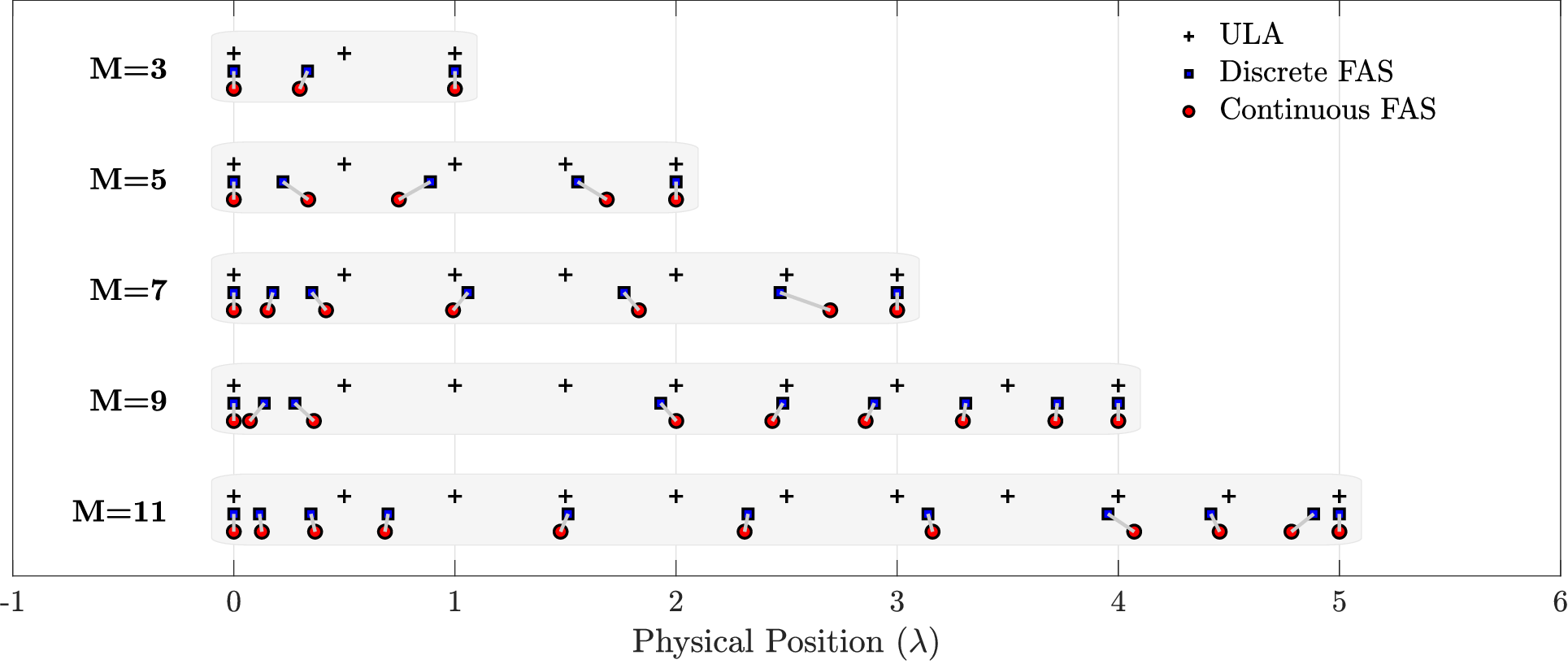}
\caption{Illustration of different arrays for $M\in \{3,5,7,9,11\}$.}\label{fig:optimized_Placement}
\end{figure}

\begin{figure}[t]
\centering
\subfloat[]{\includegraphics[width=0.495\linewidth]{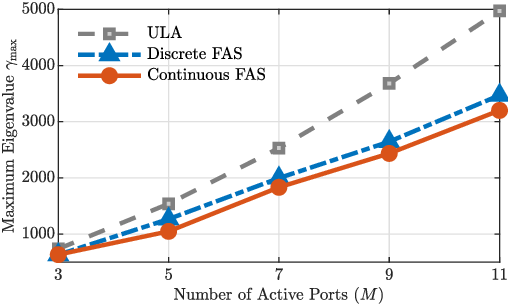}}\hspace{0pt}%
\subfloat[]{\includegraphics[width=0.495\linewidth]{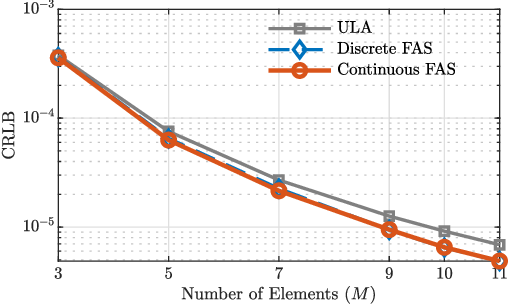}}%
\caption{Maximum eigenvalue of the sensing codebook $\boldsymbol{A}$ and CRB of different arrays: {\em (a)} $M\in\{3,5,7,9,11\}$; {\em (b)} $\mathrm{SNR}=10$ dB, target angle $\theta=15^\circ$, snapshots $T=100$.}\label{fig:CRB_eig}
\end{figure}

\begin{figure}[t]
\centering
\subfloat[]{\includegraphics[width=0.495\linewidth]{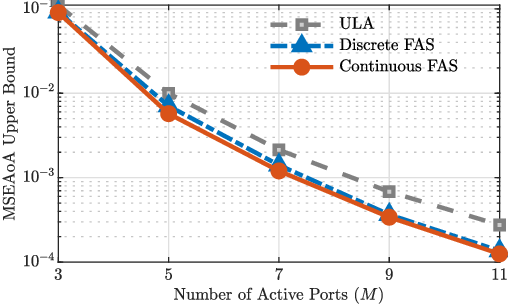}}\hspace{0pt}%
\subfloat[]{\includegraphics[width=0.495\linewidth]{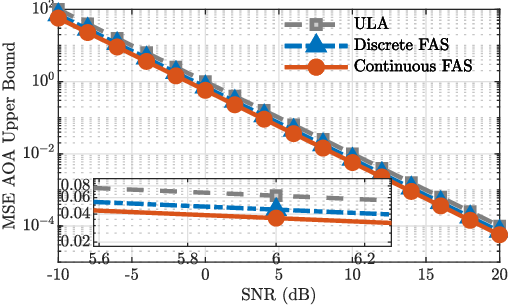}}%
\caption{AoAMSE upper bound of different arrays: {\em (a)} $\mathrm{SNR}=10$ dB, $M\in\{3,5,7,9,11\}$; {\em (b)} $\mathrm{SNR}\in\{-10,-8,\ldots,20\}$ dB, $M=5$.}\label{fig:MSE_Upper}
\end{figure}

The results in Fig.~\ref{fig:d_min} explicitly show the agreement between Monte Carlo samples and the predicted distributions, validating the fundamental limits of the minimum port placement constraint in Proposition~\ref{prop.1}.
	
In Fig.~\ref{fig:convergence}, we see that the proposed gradient-based algorithm converges under various settings, and the optimized objective value is significantly lower than that of the conventional ULA. The small step size $\alpha=5\times 10^{-4}$ explains the requirement of several hundred iterations for convergence. Nevertheless, the array design is performed offline and requires only a one-time computation. Different antenna placements are also illustrated in Fig.~\ref{fig:optimized_Placement} for clarity.

Compared with the ULA, Fig.~\ref{fig:CRB_eig} indicates that FAA exhibits a much improved sensing matrix structure with a significantly reduced maximum eigenvalue. In terms of CRB, a reduction of about $30\%$ is observed at $M=11$. The continuous and discrete cases yield similar CRB performance, and the performance gap between FAA and ULA widens as $M$ increases.

Finally, Fig.~\ref{fig:MSE_Upper} illustrates that with increasing aperture size, the AoAMSE gap between FAS and FAA becomes more pronounced. For $M=5$, an average AoAMSE reduction of $42.5\%$ is achieved across different SNR regimes.

\vspace{-2mm}
\section{Conclusion}
In this letter, we characterized the expected minimum placement gap for FAA via an accurate closed-form PDF derivation. A closed-form CRB applicable to general finite-aperture array designs was further derived. Beyond the theoretical analysis, we devised a practical array optimization algorithm that substantially reduces both the CRB and the AoAMSE under a fixed aperture. Numerical results confirmed the superiority of fluid antennas over conventional ULAs for finite-aperture array design. How to realize practical design algorithm with loose initialization would be a crucial future direction.

\end{document}